\RequirePackage[counterclockwise,figuresright]{rotating}
\documentclass[sigconf,dvipsnames]{acmart}
\usepackage[utf8]{inputenc}
\usepackage[ngerman,main=english]{babel}
\usepackage{microtype}
\usepackage{ellipsis}
\addto\extrasenglish{\languageshorthands{ngerman}\useshorthands{"}}
\usepackage[autostyle]{csquotes}
\usepackage{amsmath,amsfonts}
\usepackage{tabularx}
\usepackage{booktabs}
\usepackage{multirow}
\usepackage{makecell}
\usepackage{graphicx}
\graphicspath{{fig/}}
\usepackage{adjustbox}
\usepackage{dblfloatfix}
\usepackage{tikz}
\usepackage{tikzpagenodes}
\usetikzlibrary{
	arrows,
	arrows.meta,
	babel,
	backgrounds,
	calc,
	fit,
	positioning,
	shapes,
	shadows,
	patterns
}
\usepackage{tikz-layers}
\usepackage{enumitem}
\usepackage{calc}
\usepackage{paralist}

\newcommand{\ie}{i.\,e.\@}
\newcommand{\eg}{e.\,g.\@}

\newcommand{\wrt}{w.\,r.\,t.\@}
\newcommand{\resp}{respectively\@}
\newcommand{\totalnumsys}{64}
\makeatletter
\AtBeginEnvironment{sidewaysfigure}{\let\@vspace\@vspace@orig
  \let\@vspacer\@vspacer@orig}
\AtBeginEnvironment{sidewaysfigure*}{\let\@vspace\@vspace@orig
  \let\@vspacer\@vspacer@orig}
\makeatother
\setcopyright{none}
\renewcommand\footnotetextcopyrightpermission[1]{}

\begin{document}

\pagestyle{empty}

\title{Management of Machine Learning Lifecycle Artifacts:\texorpdfstring{\\}{}
	A Survey}

\author{Marius Schlegel}
\orcid{0000-0001-6596-2823}
\affiliation{
	\institution{TU Ilmenau}
	\city{Ilmenau}
	\country{Germany}
}
\email{marius.schlegel@tu-ilmenau.de}

\author{Kai-Uwe Sattler}
\orcid{0000-0003-1608-7721}
\affiliation{
	\institution{TU Ilmenau}
	\city{Ilmenau}
	\country{Germany}
}
\email{kus@tu-ilmenau.de}

\renewcommand{\shortauthors}{Schlegel et al.}

\begin{abstract}
	The explorative and iterative nature of developing and operating machine learning (ML) applications leads to a variety of artifacts, such as datasets, features, models, hyperparameters, metrics, software, configurations, and logs. In order to enable comparability, reproducibility, and traceability of these artifacts across the ML lifecycle steps and iterations, systems and tools have been developed to support their collection, storage, and management. It is often not obvious what precise functional scope such systems offer so that the comparison and the estimation of synergy effects between candidates are quite challenging. In this paper, we aim to give an overview of systems and platforms which support the management of ML lifecycle artifacts. Based on a systematic literature review, we derive assessment criteria and apply them to a representative selection of more than 60 systems and platforms.
\end{abstract}


\maketitle


\section{Introduction}
\label{sec:intro}

Machine learning (ML) approaches are well established in a wide range of application domains. In contrast to engineering traditional software, the development of ML systems is different: data and feature preparation, model development, and model operation tasks are integrated into a unified lifecycle which is often iterated several times~\cite{Chaoji16a,Amershi19a,Google22c}. Although there are systems and tools that provide support for a broad range of tasks within the ML lifecycle, such as data cleaning and labeling, feature engineering, model design and training, experiment management, hyperparameter optimization, and orchestration, achieving comparability, traceability, and reproducibility of model and data artifacts across all lifecycle steps and iterations is still challenging.

To meet these requirements, it is necessary to capture the input and output artifacts of each lifecycle step and iteration. That includes model artifacts and data"=related artifacts, such as datasets, labels, and features. Reproducibility also requires capturing software"=related artifacts, such as code, configurations, and environmental dependencies. By additionally considering metadata, such as model parameters, hyperparameters, quality metrics, and execution statistics, comparability of artifacts is enabled.

Since the manual management of ML artifacts is simply not efficient, systems and platforms provide support for the systematic collection, storage, and management of ML lifecycle artifacts, which we collectively referred to as \emph{ML artifact management systems} (ML AMSs)\footnote{Whenever we use just \enquote{AMS}, we refer to an ML AMS.}. Since ML AMSs are often integrated into general ML development platforms or frameworks for a subset of the ML lifecycle tasks, it is typically not obvious what the precise functional and non"=functional scope of an AMS is, how an AMS compares to others, and to what extent possible synergy effects can be exploited through tool"=chaining.

The objective of this paper is to provide a comprehensive overview of AMSs from academia and industry. We address the following research questions: (RQ1)~What are criteria to describe, assess, and compare AMSs? (RQ2)~Which AMSs exist in academia and industry, and what are their functional and non"=functional properties according to the assessment criteria? To answer these questions, we conducted a systematic literature review.

The paper is organized as follows: §\,\ref{sec:relwork} gives an overview of related work. §\,\ref{sec:artifactmanagement} describes the ML lifecycle and concretizes the tasks of ML lifecycle management. Based on the conducted systematic literature review, §\,\ref{sec:criteria} discusses criteria for assessing AMSs \wrt{} their functional and non"=functional scope of features. §\,\ref{sec:assessment} applies the criteria to the \totalnumsys{} identified AMSs and discusses the results.


\section{Related Work}
\label{sec:relwork}

In recent years, both academia and industry have produced a variety of systems that provide artifact collection and management support for individual steps of ML lifecycles \cite{Miao17b,Schelter17a,Agrawal19a,Brumbaugh19a,Hummer19a,Chen20a,Ferenc20a,Barrak21a,Gharibi21a}. Authors often compare with related works in the scope of the particular system, which, however, does not enable the comparability with a broad range of systems and criteria.

This problem has been addressed by a few surveys \cite{Isdahl19a,Weissgerber19a,Idowu21a}. In the context of reproducibility of empirical results, Isdahl et al.~\cite{Isdahl19a} have investigated what support is provided by existing experiment management systems. However, these systems cover only a subset of the ML lifecycle. Weißgerber et al.~\cite{Weissgerber19a} develop an open"=science"=centered process model for ML research as a common ground and investigate 40~ML platforms and tools. However, the authors analyze only 11~platforms \wrt{} ML workflow support capabilities and their properties.

In contrast to the aforementioned studies and surveys, Idowu et al.~\cite{Idowu21a} adopt a more fine"=grained understanding of artifacts and system capabilities which is most closely to our work. Based on a selection of 17~experiment management systems and tools, the authors develop a feature model for assessing their capabilities. Although this survey shows parallels to our work, the authors only consider a limited selection of systems which is, again, only focused on the area of experiment tracking and management.


\section{Machine Learning Lifecycle Artifact Management}
\label{sec:artifactmanagement}

In this section, we discuss the steps of ML lifecycles based on typical ML workflows (§\,\ref{sec:artifactmanagement:mllifecycle}), derive the tasks of ML artifact management, and outline the support ML AMSs should provide (§\,\ref{sec:artifactmanagement:artifactsandmanagement}).

\subsection{ML Lifecycle}
\label{sec:artifactmanagement:mllifecycle}

\begin{figure}[!b]
	\vspace{-\baselineskip}
	\centering
	\hspace*{-9pt}
	\begin{tikzpicture}
		\tikzstyle{every node}=[
		node font=\small,
		minimum height=10mm,
		minimum width=28mm,
		inner sep=1mm,
		align=center,
		node distance=5mm and 8mm,
		]
		\tikzstyle{nod}=[
		fill=white,
		drop shadow,
		draw,
		]
		\def\stagedist{28pt}
		\node [nod] (datacoll)                                                  {Data Collection\\and Selection};
		\node [nod] (dataclean)      [right=of datacoll]                        {Data Cleaning};
		\node       (abovedatacoll)  [above=\stagedist of datacoll]             {};
		\node       (abovedataclean) [above=\stagedist of dataclean]            {};
		\node [nod] (modreq)         at ($(abovedatacoll)!.5!(abovedataclean)$) {Model Requirements\\Analysis};
		\node [nod] (datalab)        [below=of dataclean]                       {Data Labeling};
		\node [nod] (feateng)        [below=of datacoll]                        {Feature Engineering\\and Selection};
		\node [nod] (moddesign)      [below=\stagedist of feateng]              {Model Design};
		\node [nod] (modtrain)       [below=\stagedist of datalab]              {Model Training};
		\node [nod] (modeval)        [below=of modtrain]                        {Model Evaluation};
		\node [nod] (modoptim)       [below=of moddesign]                       {Model Optimization};
		\node [nod] (moddeploy)      [below=\stagedist of modoptim]             {Model Deployment};
		\node [nod] (modmon)         [below=\stagedist of modeval]              {Model Monitoring};
		\draw[->,>=latex,semithick] (datacoll)  edge (dataclean);
		\draw[->,>=latex,semithick] (dataclean) edge (datalab);
		\draw[->,>=latex,semithick] (datalab)   edge (feateng);
		\draw[->,>=latex,semithick] (moddesign) edge (modtrain);
		\draw[->,>=latex,semithick] (modtrain)  edge (modeval);
		\draw[->,>=latex,semithick] (modeval)   edge (modoptim);
		\draw[->,>=latex,semithick] (modoptim)  edge (modtrain);
		\draw[->,>=latex,semithick] (modoptim)  edge (moddesign);
		\draw[->,>=latex,semithick] (moddeploy) edge (modmon);
		\tikzstyle{outbox}=[
		shape=rectangle,
		draw,
		dotted,
		thick,
		rounded corners=1.5mm,
		inner sep=5pt,
		black!70
		]
		\begin{scope}[on behind layer]
			\node[fit={(abovedatacoll)(abovedataclean)},outbox,fill=yellow!10,
				label={[name=l,rotate=90,anchor=south,yshift=2pt,font=\itshape]left:Requirements\\Stage}
			] (reqstage) {};
			\node[fit={(datacoll)(dataclean)(datalab)(feateng)},outbox,fill=red!10,
				label={[name=l,rotate=90,anchor=south,yshift=2pt,font=\itshape]left:Data-oriented\\Stage}
			] (datastage) {};
			\node[fit={(moddesign)(modtrain)(modeval)(modoptim)},outbox,fill=cyan!10,
				label={[name=l,rotate=90,anchor=south,yshift=2pt,font=\itshape]left:Model-oriented\\Stage}
			] (modelstage) {};
			\node[fit={(moddeploy)(modmon)},outbox,fill=ForestGreen!10,
				label={[name=l,rotate=90,anchor=south,yshift=2pt,font=\itshape]left:Operations\\Stage}
			] (opstage) {};
		\end{scope}
		\draw[->,>=latex,line width=2pt] (reqstage)   edge (datastage);
		\draw[->,>=latex,line width=2pt] (datastage)  edge (modelstage);
		\draw[->,>=latex,line width=2pt] (modelstage) edge (opstage);
		\coordinate (modelstage2) at ($(modelstage.east)-(-8pt,-15pt)$) {};
		\draw[->,>=latex,semithick,line width=2pt] (modelstage.east) arc (-90:0:8pt) -- (modelstage2);
		\coordinate (opstage2) at ($(opstage.east)-(-8pt,-15pt)$) {};
		\draw[->,>=latex,semithick,line width=2pt] (opstage.east) arc (-90:0:8pt) -- (opstage2);
	\end{tikzpicture}
	\vspace{-2\baselineskip}
	\caption{Typical ML lifecycle.}
	\label{fig:mllifecycle}
	\Description[Typical ML lifecycle]{Typical ML lifecycle}
\end{figure}
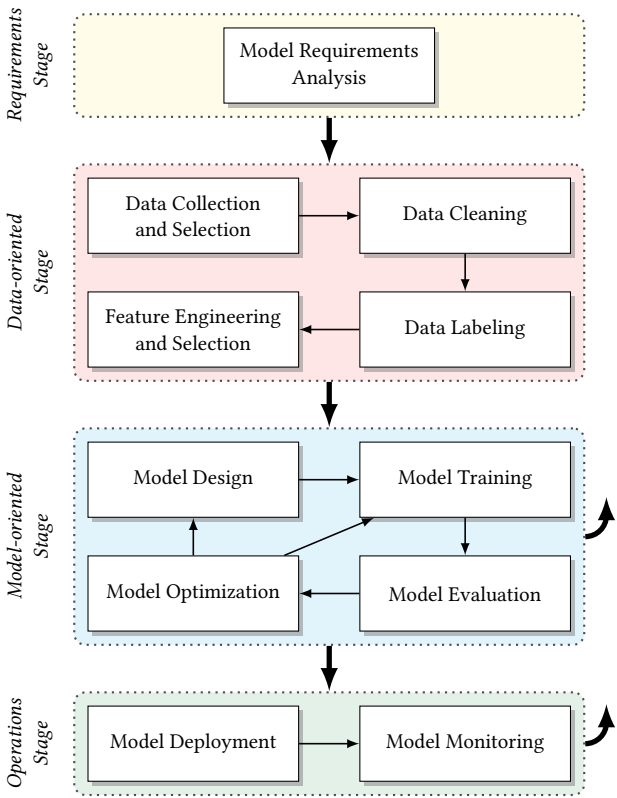

In contrast to traditional software engineering, the development of ML-powered applications is more iterative and explorative. Thus, developers have adapted their processes and practices for ML: Following methodologies in the context of data science, data analytics and data mining, such as TDSP \cite{Microsoft20a}, KDD \cite{Fayyad96a}, CRISP"=DM \cite{Colin00a,Wirt00a}, or ASUM"=DM \cite{IBM16a}, workflows specialized for ML have been established \cite{Chaoji16a,Amershi19a,Google22c,Salvaris18a}. Despite minor differences, ML workflows contain both data"=centric and model"=centric steps and often multiple feedback loops among the different stages, which leads to a \emph{lifecycle}. Fig.~\ref{fig:mllifecycle} depicts a common view on the ML lifecycle.

The ML lifecycle consists of four stages: Requirements Stage, Data"=oriented Stage, Model"=oriented Stage, and Operations Stage. Starting with the Requirements Stage, the requirements for the model to be developed are derived based on the application requirements \cite{Vogelsang19a}. This stage is dedicated to three major decisions: (1.)~to decide which functionality and interfaces to realize, (2.)~to decide which types of models are best suited for the given problem, and (3.) to decide which types of data to work on.

The Data-oriented Stage starts with the Data Collection and Selection step. Datasets, either internal or publicly available, are searched, or individual ones are collected and the data most suitable for the subsequent steps is selected (\eg{} dependent on data quality, bias, etc.). By using available generic datasets, models may be (pre-)""trained (\eg{} Image"-Net for object recognition), and later, by using transfer learning \cite{Bozinovski20a,Ng16a} along with more specific data, trained to a more specific model. Then, in the Data Cleaning step, datasets are prepared, removing inaccurate or noisy records. As required by most of the supervised learning techniques to be able to induce a model, data labeling is used to assign a ground truth label to each dataset record. Subsequently, feature engineering and selection is performed to extract and select features for ML models. For some models, such as convolutional neural networks, this step is directly intertwined with model training.

The Model-oriented Stage starts with the Model Design step. Often, existing model designs and neural network architectures are used and tailored towards specific requirements. During model training, the selected models are trained on the collected and preprocessed datasets using the selected features and their respective labels. Subsequently, in the Model Evaluation step, developers evaluate a trained model on test datasets using predefined metrics, such as accuracy or F1-score. In critical application domains, this step also includes extensive human evaluation. The subsequent Model Optimization step is used to fine"=tune the model, especially its hyperparameters. In the context of the model development steps, we refer to an experiment as a sequence of model development activities that result in a trained model but do not include cycles to previous steps.

Finally, in the Operations Stage, the model is distributed to the target systems and devices, either as an on"=demand (online) service or in batch (offline) mode (Model Deployment), as well as continuously monitored for metrics and errors during execution and use (Model Monitoring). In particular, CI/""CD practices from software engineering are adapted.

As illustrated by Fig.~\ref{fig:mllifecycle}, multiple feedback loops from steps of the Model"=oriented Stage or the Operations Stage to any step before may be triggered by insufficient accuracy or new data. Moreover, Sculley et al.~\cite{Sculley15a} point out, that the model development often takes only a fraction of the time required to complete ML projects. Usually, a large amount of tooling and infrastructure is required to support data extract, transform, and load (ETL) pipelines, efficient training and inference, reproducible experiments, versioning of datasets and models, model analysis, and model monitoring at scale. The creation and management of services and tools can ultimately account for a large portion of the workload of ML engineers, researchers, and data scientists.

\subsection{Management of ML Lifecycle Artifacts}
\label{sec:artifactmanagement:artifactsandmanagement}

Within the steps of the ML lifecycle, a variety of artifacts is created, used, and modified: Datasets, labels and annotations, and feature sets are inputs and outputs of steps in the Data"=oriented Stage. Moreover, data processing source code, logs, and environmental dependencies are created and/""or used. In the Model"=oriented Stage, results from the Data"=oriented Stage are used to develop and train models. In addition, metadata such as parameters, hyperparameters, and captured metrics as well as model processing source code, logs, and environment dependencies are artifacts that are created and/""or used in this stage. The Operations Stage requires trained models and corresponding dependencies such as libraries and runtimes (\eg{} via Docker container), uses model deployment and monitoring source code which is typically wrapped into a web service with a REST API for on"=demand (online) service or scheduled for batch (offline) execution, and captures execution logs and statistics. To achieve comparability, traceability, and reproducibility of produced data and model artifacts across multiple lifecycle iterations, it is essential to also capture metadata artifacts that can be easily inspected afterwards (\eg{} model parameters, hyperparameters, lineage traces, performance metrics) as well as software artifacts.

Manual management of artifacts is simply not efficient due to the complexity and the required time. To meet the above requirements, it is necessary to systematically capture any input and output artifacts and to provide them via appropriate interfaces. \emph{ML artifact management} includes any methods and tools for managing ML artifacts that are created and used in the development, deployment, and operation of ML"=based systems. Systems supporting ML artifact management, collectively referred to as \emph{ML artifact management systems} (ML AMSs), provide the functionality and interfaces to adequately record, store, and manage ML lifecycle artifacts.


\section{Assessment Criteria}
\label{sec:criteria}

\begin{table*}[!t]
	\vspace*{.45\baselineskip}
	\centering
	\small
	\newlength\colonewidth
	\settowidth{\colonewidth}{\textbf{Collection \&\hspace{1pt}}}
	\setlength{\extrarowheight}{1pt}
	\begin{tabularx}{.85\textwidth}{@{}p{\colonewidth}X@{}}
		\toprule
		\textbf{Category} & \textbf{Criteria and Subcriteria}                                                                                                                                      \\
		\midrule
		\multirowcell{5}[0pt][l]{\textbf{Lifecycle}                                                                                                                                                \\\textbf{Integration}}
		                  & \textit{Requirements Stage} [Model Requirements Analysis]                                                                                                              \\
		                  & \textit{Data-oriented Stage} [Data Collection \& Selection, Data Preparation \& Cleaning, Data Labeling, Feature Engineering \& Selection]                             \\
		                  & \textit{Model-oriented Stage} [Model Design, Model Training, Model Evaluation, Model Optimization]                                                                     \\
		                  & \textit{Operations Stage} [Model Deployment, Model Monitoring]                                                                                                         \\
		\midrule
		\multirowcell{5}[0pt][l]{\textbf{Artifact}                                                                                                                                                 \\\textbf{Support}}
		                  & \textit{Data-related Artifacts} [Dataset, Annotations \& Labels, Features]                                                                                             \\
		                  & \textit{Model Artifacts} [Model]                                                                                                                                       \\
		                  & \textit{Metadata Artifacts} [Identification, Model Parameters, Model Hyperparameters, Model Metrics, Experiments \& Projects, Pipelines, Execution Logs \& Statistics] \\
		                  & \textit{Software Artifacts} [Source Code, Notebooks, Configurations, Environment]                                                                                      \\
		\midrule
		\multirowcell{4}[0pt][l]{\textbf{Operations}}
		                  & \textit{Logging \& Versioning} [Log/Capture, Commit, Revert/Rollback]                                                                                                  \\
		                  & \textit{Exploration} [Query, Compare, Lineage, Provenance, Visualize]                                                                                                  \\
		                  & \textit{Management} [Modify, Delete, Execute \& Run, Deploy]                                                                                                           \\
		                  & \textit{Collaboration} [Export \& Import, Share]                                                                                                                       \\
		\midrule
		\multirowcell{3}[0pt][l]{\textbf{Collection \&}                                                                                                                                            \\\textbf{Storage}}
		                  & \textit{Collection Automation}  [Intrusive, Non-intrusive]                                                                                                             \\
		                  & \textit{Storage Type} [Filesystem, Database, Object/BLOB Storage, Repository]                                                                                          \\
		                  & \textit{Versioning} [Repository, Snapshot]                                                                                                                             \\
		\midrule
		\multirowcell{2}[0pt][l]{\textbf{Interfaces \&}                                                                                                                                            \\\textbf{Integration}}
		                  & \textit{Interface} [API/SDK, CLI, Web UI]                                                                                                                              \\
		                  & \textit{Language Support \& Integration} [Languages, Frameworks, Notebook]                                                                                             \\
		\midrule
		\multirowcell{2}[0pt][l]{\textbf{Operation \&}                                                                                                                                             \\\textbf{Licensing}}
		                  & \textit{Operation} [Local, On-premise, Cloud]                                                                                                                          \\
		                  & \textit{License} [Free, Non-free]                                                                                                                                      \\
		\bottomrule
	\end{tabularx}
	\vspace{.9\baselineskip}
	\caption{Assessment categories, criteria, and subcriteria.}
	\label{tab:criteria}
	\vspace{-.69\baselineskip}
\end{table*}

The goal of this section is to define criteria for the description and assessment of AMSs. Based on a priori assumptions, we first list functional and non"=functional requirements. We then conduct a systematic literature review according to Kitchenham et al. \cite{Kitchenham07a}: Using well"=defined keywords, we search ACM DL, DBLP, IEEE Xplore, and Springer"-Link for academic publications as well as Google and Google Scholar for web pages, articles, white papers, technical reports, reference lists, source code repositories, and documentations. Next, we perform the publication selection based on the relevance for answering our research questions. To avoid overlooking relevant literature, we perform one iteration of \emph{backward snowballing} \cite{Wohlin14a}. Finally, we iteratively extract assessment criteria and subcriteria, criteria categories, as well as the functional and non"=functional properties of concrete systems and platforms based on concept matrices. The results are shown in Tab.~\ref{tab:criteria}, which outlines categories, criteria (italicized), subcriteria (in square brackets).

\paragraph{Lifecycle Integration}

This category describes for which parts of the ML lifecycle a system provides artifact collection and management capabilities. The four stages form the criteria, with the steps assigned to each stage forming the subcriteria (cf. §\,\ref{sec:artifactmanagement:mllifecycle}).

\paragraph{Artifact Support}

Orthogonal to the previous category, this category indicates which types of artifacts are supported and managed by an AMS. Based on the discussion in §\,\ref{sec:artifactmanagement:artifactsandmanagement}, we distinguish between the criteria \textit{Data-related}, \textit{Model}, \textit{Metadata}, and \textit{Software Artifacts}.

The criteria \textit{Data"=related Artifacts} and \textit{Model Artifacts} represent core resources that are either input, output, or both for a lifecycle step. \textit{Data"=related Artifacts} are datasets (used for training, validation, and testing), annotations and labels, and features (cf. corresponding subcriteria). \textit{Model Artifacts} are represented by trained models (subcriterion \textit{Model}).

The criteria \textit{Metadata Artifacts} and \textit{Software Artifacts} represent the corresponding artifact types, that enable the reproducibility and traceability of individual ML lifecycle steps and their results. The criterion \textit{Metadata Artifacts} covers different types of metadata:
(i)~identification metadata (\eg{} identifier, name, type of dataset or model, association with groups, experiments, pipelines, etc.);
(ii)~data-related metadata;
(iii)~model-related metadata, such as
inspectable model parameters (\eg{} weights and biases),
model hyperparameters (\eg{} number of hidden layers, learning rate, batch size, or dropout), and
model quality \& performance metrics (\eg{} accuracy, F1-score, or AUC score);
(iv)~experiments and projects, which are abstractions to capture data processing or model training runs and to group related artifacts in a reproducible and comparable way;
(v)~pipelines, which are abstractions to execute entire ML workflows in an automated fashion and relates the input and output artifacts required per step as well as the glue code required for processing;
(vi)~execution"=related logs \& statistics.

The criterion \textit{Software Artifacts} comprises source code and notebooks, \eg{} for data processing, experimentation and model training, and serving, as well as configurations and execution"=related environment dependencies and containers, \eg{} Conda environments, Docker containers, or virtual machines.

\paragraph{Operations}

This category indicates the operations provided by an AMS for handling and managing ML artifacts. It comprises the criteria \textit{Logging \& Versioning}, \textit{Exploration}, \textit{Management}, and \textit{Collaboration}.

The criterion \textit{Logging \& Versioning} represents any operations that enable logging or capturing single artifacts (subcriterion \textit{Log/""Capture}), creating checkpoints of a project or an experiment comprising several artifacts (subcriterion \textit{Commit}), and reverting or rolling back to an earlier committed or snapshot version (subcriterion \textit{Revert/""Rollback}).

The criterion \textit{Exploration} includes any operations that help to gain concrete insights into the results of data processing pipelines, experiments, model training results, or monitoring statistics. These operations are differentiated by the subcriteria \textit{Query}, \textit{Compare}, \textit{Lineage}, \textit{Provenance}, and \textit{Visualize}.
\textit{Query} operations may be represented by simple searching and listing functionality, more advanced filtering functionality (\eg{} based on model performance metrics), or a comprehensive query language.
\textit{Compare} indicates the presence of operations for the comparison between two or more versions of artifacts. In terms of model artifacts, this operation may be used to select the most promising model from a set of candidates (\emph{model selection}), either in model training and development~\cite{Raschka18a} or in model serving (\eg{} best performing predictor)~\cite{Crankshaw17a}.
\textit{Lineage} represents any operations for tracing the lineage of artifacts, \ie{} which input artifacts led to which output artifacts, and thus provide information about the history of a model, dataset, or project.
\textit{Provenance} represents any operations, which in addition provide information about which concrete transformations and processes converted inputs into an output.
\textit{Visualize} indicates the presence of functionality for graphical representation of model architectures, pipelines, model metrics, or experimentation results.

The criterion \textit{Management} characterizes operations for handling and using stored artifacts. The subcriteria \textit{Modify} and \textit{Delete} indicate operations for modifying or deleting logged and already stored artifacts. \textit{Execute \& Run} comprises operations that are interfaces for the execution and orchestration of data processing or model training experiments and pipelines. The subcriterion \textit{Deploy} refers to deployment operations for offline (batch) and online (on"=demand) model serving.

The criterion \textit{Collaboration} indicates the presence of operations for collaboration which enable simple export/""import functionality (subcriterion \textit{Export \& Import}) as well as sharing of artifacts among internal company and team members or publishing of artifacts for external instances (subcriterion \textit{Share}).

\paragraph{Collection \& Storage}

This category describes the model for artifact collection and storage based on the criteria \textit{Collection Automation}, \textit{Storage Type}, and \textit{Versioning}.

The criterion \textit{Collection Automation} represents the degree of manual effort required to collect and capture ML artifacts. The collection of artifacts is either intrusive, which requires engineers to explicitly add instructions or API calls within the source code, non"=intrusive, which means that no explicit manual actions are required and the collection is performed automatically, or both.

\textit{Storage Type} describes the type of storage used and supported by an AMS. We identified the subcriteria \textit{Filesystem}, \textit{Database}, \textit{Object/""BLOB Storage} and \textit{Repository}. An AMS can support multiple types of storage, and also hybrid variants are possible.

While local filesystems are often the means of choice to store artifacts for small and manageable experiments, such as smaller textual datasets (\eg{} in \texttt{.csv} or \texttt{.parqet} files) or metadata (\eg{} in \texttt{.yaml} files), distributed file systems (\eg{} HDFS) are rather used for larger projects and permanently scalable solutions. These are suitable for both large numbers and large files, as is the case for image, video, and text datasets as well as trained models in serialized file formats (\eg{} \texttt{.pb}, \texttt{.onnx}, \texttt{.pkl}, \texttt{.pt}, or \texttt{.pmml}). Often, these are also stored on separate storage servers or clusters, and provided with an API to fulfill availability and replication requirements.

Databases are established for storing a wide range of different types of data. In the context of artifact management, large tabular, sequential, and text datasets may be stored in (object-)relational or time"=series databases, metadata in relational databases, and logs in key-value databases. While modern and widely used DBMSs (\eg{} PostgreSQL \cite{PostgreSQL22a}) provide BLOB data types, these often have limitations regarding maximum file sizes, which is why also object/""BLOB stores are often used.

Repositories are version-controlled and typically suited best for source code and text. It is also possible to add version control on top of other storage types, so that the storage of large files in a distributed file system is combined with version control. For example, Git LFS (Large File Storage) replaces large files such as image, audio, and video datasets, with text pointers inside Git, while storing the file contents on a remote file server. This preserves and accelerates typical workflows, while enabling versioning of large amounts of data or large files.

The criterion \textit{Versioning} characterizes the way versioning of artifacts is done. Either versioning is delegated to a traditional version control system (\eg{} Git or Mercurial) and managed by means of a repository (see also corresponding storage type). A repository contains several to all artifacts associated with a project. In contrast, versioning may be done by following a snapshot"=based approach: Snapshots are created manually for each individual artifact and possibly independently of other artifacts.

\paragraph{Interfaces \& Integration}

This category characterizes an AMS's user interfaces and integration capabilities.
The criterion \textit{Interface} states the type of provided interfaces that may be based on an API (\eg{} a REST interface) or a higher"=level SDK, based on a command line interface (CLI), or based on a web application.
\textit{Language Support \& Integration} distinguishes between the integration into programming languages (\eg{} into Python via provided libraries), integration into well"=known frameworks providing functional integration for the steps of the ML lifecycle (\eg{} data processing with Apache Spark \cite{Spark22a}, model training with TensorFlow \cite{TensorFlow22a}, or model orchestration with Metaflow \cite{Metaflow22a}) and notebook support (\eg{} Jupyter \cite{Jupyter22a}, Apache Zeppelin \cite{Zeppelin22a}, or TensorBoard \cite{TensorBoard22a})

\paragraph{Operation \& Licensing}

The last category covers two non"=functional, usage"=related criteria \textit{Operation} and \textit{License}.
The criterion \textit{Operation} defines whether the operation of a system or tool is local (\eg{} the case for Python libraries, subcriterion \textit{Local}), on"=premise (\eg{} the case for server"=based systems, sub"-criterion \textit{On"=premise}) or by a dedicated cloud provider (\eg{} for hosted services, sub"-criterion \textit{Cloud}).
The criterion \textit{License} describes the type of license, which is either classified as free (\eg{} public domain, permissive, or copyleft licenses) or non"=free (\eg{} non"=commercial or proprietary licenses), and which may be further concretized by the concrete license.


\section{Assessment}
\label{sec:assessment}

\begin{sidewaysfigure*}[!p]
	\begin{adjustbox}{clip,trim=8pt 6pt 8pt 40pt,max width=\textwidth}
		\input{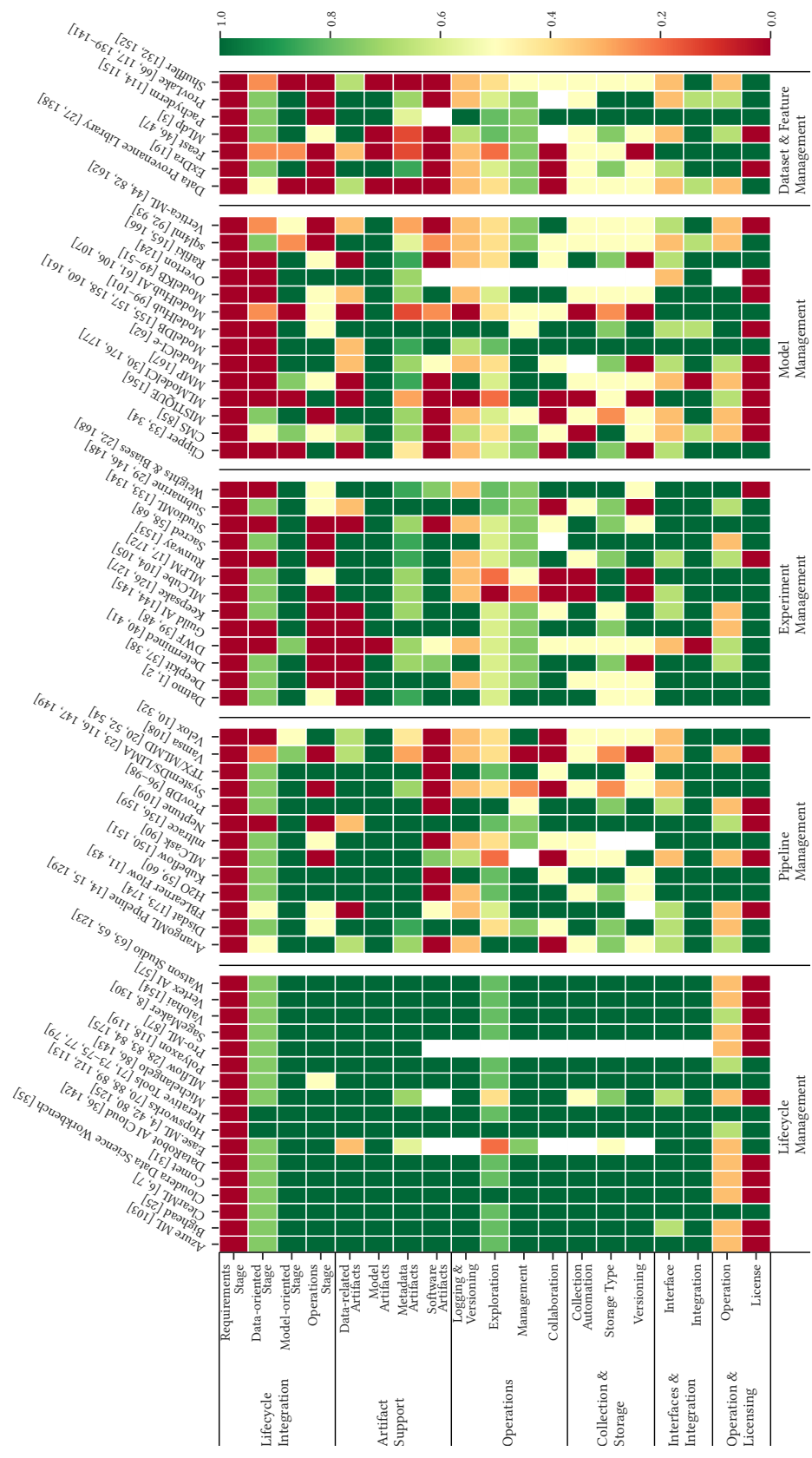}
	\end{adjustbox}
	\caption{%
		The heatmap visualizes the assessment results. For each criterion, the number of subcriteria fulfilled/""present was determined, related to the total number of subcriteria, and normalized to the value range [0,\,1].
		The degree of fulfillment/""presence of a criterion (y-axis) by the investigated systems (x-axis) is represented by the hue of a cell ranging from dark red (\ie{} not fulfilled/""present) to dark green (\ie{} completely fulfilled/""present).
		If the fulfillment/""presence of all subcriteria of a criterion is unclear or not exactly known, the corresponding cell's hue is white.
		The last criterion \enquote{License} is an exception due to its binary character: the hue is either dark red (\enquote{non-free}) or dark green (\enquote{free}).}
	\label{fig:assessmentresults}
\end{sidewaysfigure*}

This section presents the assessment of concrete ML AMSs. As part of the systematic literature review (see §\,\ref{sec:criteria}), we identified a total of \totalnumsys{}~systems and platforms from research and industry. We assessed these based on our criteria and subcriteria (cf. §\,\ref{sec:criteria}). An additional result of this assessment is the derivation of the classes which aim to group systems with high similarity regarding lifecycle integration, artifact support, and functionality. Fig.~\ref{fig:assessmentresults} visualizes the results at criteria level (depending on a criterion's semantic, we consider either its fulfillment or presence) and the AMS classes.
The following two subsections discuss the assessment results on these orthogonal dimensions:
Based on the derived classes, we first provide an overview of the identified AMSs and their core characteristics (§\,\ref{sec:assessment:classdiscussion}). Subsequently, we discuss to what extent the criteria and subcriteria are fulfilled by the systems within their classes (§\,\ref{sec:assessment:criteriadiscussion}).

\subsection{Discussion Along Classes}
\label{sec:assessment:classdiscussion}

A general observation is that the focus of ML"=supporting systems is often not obvious: The boundaries are blurred between systems that purely provide functionality for the development of ML"=based systems to those that purely focus on the management, storage, and deployment of ML artifacts (and typically complement the former). Therefore, we classified the assessed systems based on the characteristics of the criteria within the categories Lifecycle Integration, Artifact Support, and Operations into five classes: Lifecycle Management, Pipeline Management, Experiment Management, Model Management, and Dataset \& Feature Management.

\paragraph{Lifecycle Management}

This class includes systems and platforms that focus on the entire ML lifecycle and, typically beyond broad functional support for the different tasks within the ML lifecycle, provide artifact management capabilities for nearly all lifecycle steps.

Microsoft \emph{Azure ML} \cite{Microsoft22a}, Amazon \emph{SageMaker} \cite{Schelter17a,Amazon22a}, Google \emph{Vertex AI} \cite{VertexAI22a}, IBM \emph{Watson Studio} \cite{Hummer19a,Rausch20a,IBM22a}, \emph{Comet} \cite{CometML22a}, \emph{Data"-Robot AI Cloud Platform} \cite{Sridhar18a,DataRobot22a}, and \emph{Cloudera Data Science Workbench} \cite{Cloudera22a} are \enquote{all"=in"=one} ML as a service (MLaaS) platforms which are comparable in objective and functional scope. Although the direct integration into a provider's cloud infrastructure usually offers rich processing possibilities, such as the scaling of computing and memory resources, the usage is subject to the fees and pricing of the provider. Furthermore, the usage of MLaaS may not be possible in certain application scenarios due to data protection requirements and legal regulations.

In contrast, open"=source AMSs, such as \emph{MLflow} \cite{Zaharia18a,Chen20a,MLflow22a,MLflow22b}, \emph{ClearML} \cite{ClearML22a,ClearML22b}, \emph{Polyaxon} \cite{Polyaxon22a,Polyaxon22b}, and \emph{Hopsworks} \cite{Ismail17a,Ormenisan20a,Ormenisan20b,Hopsworks22a,Hopsworks22b}, can be deployed both in the cloud and on"=premise.
MLflow focuses on capturing, storing, managing, and deploying ML artifacts: MLflow Tracking is an API for logging experiment runs, including code and data dependencies, via automatic or manual instrumenting application code. These runs can be viewed, compared, and search through an API or the UI. MLflow Models are a convention for packaging models and their dependencies, that is compatible with diverse serving environments. MLflow Projects provides a standard format for packaging reusable and reproducible project code. MLflow Model Registry is a hub for storing models and managing their deployment lifecycles.
ClearML and Polyaxon also aim at simplifying the entire ML lifecycle. Both provide a comparable range of functions, plus model monitoring and resource management.
The Hopsworks platform also has a comparable range of functions but additionally includes big data and GPU support for highly scalable learning, HDFS extended by a distributed hierarchical metadata service using a NewSQL database (\emph{HopsFS}), Github"=like project management, and an integrated feature store.

\emph{Valohai}~\cite{Valohai22a} is a platform for managing and versioning ML pipelines from data extraction to model deployment. Its objective is comparable to the previous AMSs. The three layers of the platform, (Web) Application Layer, Computer Layer, and Data Layer, can be flexibly deployed in the cloud (\eg{} Valohai or own AWS account) or on"=premise.

In relation to the previously described AMSs, \emph{DVC}~\cite{Barrak21a,DVC22a,DVC22b} is a complementary ML artifact management tool for ML pipe"-lines. DVC's versioning is built upon Git and provides experiment branching semantics, push and pull processing of bundles of models, data, and code, as well as automatic metric tracking. Recently, the company behind, Iterative, built a tool ecosystem around DVC to achieve ML lifecycle management~\cite{Iterative22a}: a library for implementing CI/CD in ML projects (CML)~\cite{CML22a,CML22b}, a library for logging ML metrics and metadata (DVCLive)~\cite{DVCLive22a}, a web application for seamless data and model management, experiment tracking, visualization, and collaboration (Studio)~\cite{Studio22a,Studio22b}, a model registry and deployment tool (MLEM)~\cite{MLEM22a}.

In contrast, \emph{Ease.ML} is an ML lifecycle management system specifically targeting non"=ML experts \cite{Aguilar21a,EaseML22a,Renggli19a,Karlas18a}. Built on top of existing data ecosystems and techniques, Ease.ML guides users step"=by"=step: Starting by automatic data ingestion and augmentation, automatic feasibility studies, data noise debugging, data acquisition, scalable multi"=tenant automatic training, continuous integration, and ending with continuous quality optimization.

Additionally, we identified proprietary and internally deployed ML lifecycle management platforms from the major tech companies Airbnb, Uber, and LinkedIn.
\emph{Bighead} is Airbnb's framework"=agnostic ML platform tailored to their use cases and environment \cite{Brumbaugh19a}. It includes a feature management framework based on the lambda architecture principle \cite{Marz11a,Marz13a}, a model development and execution management toolkit, a lifecycle management service, an offline training and inference engine, an online inference service with containerization and cloud native architecture, and a container management~tool.

\emph{Michelangelo} is Uber's ML platform which enables internal teams to build, deploy, and operate uniform and reproducible ML pipelines for applications in a microservices"=based production environment \cite{Li17a}. Michelangelo consists of a mix of open"=source systems and components built in"=house, such as a centralized feature store, a domain"=specific language (DSL) for feature selection and transformation, the distributed deep learning framework \emph{Horovod} \cite{Sergeev18a} and the model management system \emph{Gallery} \cite{Sun20a}.

LinkedIn's \emph{Pro-ML} system specifically aims to meet internal scalability requirements \cite{LinkedIn19a}. Pro"=ML supports a key set of ML lifecycle steps: data exploration and model authoring with an own DSL for feature and model representations and a central feature marketplace, real time and batch model training, model deployment, and model monitoring.

\paragraph{Pipeline Management}

ML pipelines are abstractions to enable a holistic view on data processing, model development, and model deployment workflows. This class includes any AMSs that support the management of ML pipelines and related artifacts. In contrast to the previous class, these systems typically do not comprise either support for the Operations Stage (in a few cases), the management of software artifacts, or both.

Although the pipeline management systems \emph{Velox} \cite{Crankshaw15a,Velox22a}, \emph{Vamsa} \cite{Namaki20a}, and \emph{ArangoML Pipeline} \cite{Schad21a,ArrangoML22a,ArrangoML22b} differ to some extent in their goals, design principles, and intended uses, they overall provide a comparable basic range of artifact management capabilities, with a focus on models and model metadata.
Velox provides management of training pipeline orchestration for a set of pre"-declared models, model performance evaluation, retraining as necessary, and low"=latency model inference.
Vamsa is a tool for automated provenance tracking of ML pipelines based on static analyses of Python programs.
ArangoML Pipeline is an artifact and metadata storage layer for ML pipeline lineage tracking, auditing, reproducibility, and monitoring built around Arango"-DB.

In comparison, Apache \emph{System"-DS} \cite{Boehm20a,SystemDS22a,SystemDS22b} is a declarative ML pipeline system. It provides a DSL with declarative language abstractions for different ML lifecycle tasks. High"=level scripts are compiled into hybrid execution plans of local, in"=memory CPU and GPU operations, as well as distributed operations on Spark based on a tensor data model. Based on the LIMA framework, System"-DS provides fine"=grained, multi"=level lineage tracing as well as compiler"=assisted, full and partial reuse during runtime removing redundancy at different levels of hierarchically composed ML pipelines~\cite{Phani21a}.

\emph{Mltrace} \cite{Shankar21a,Mltrace22a} and \emph{ProvDB} \cite{Miao17c,Miao18a,Miao18b} also have a strong focus on lineage and provenance tracing.
Mltrace is a Python tool for lineage and tracing of artifacts in ML pipe"-lines. It integrates into existing codebases without requiring redesigning pipelines or rewriting pipeline code.
\emph{ProvDB} is a unified provenance, model lineage, and metadata management system founded on a graph"=based provenance representation model that generalizes the W3C PROV data model. It features the Neo4j Cypher and Apache Gremlin graph query languages and two query operators for graph segmentation and summarization geared towards the characteristics of provenance information.

Furthermore, the \emph{TensorFlow Extended (TFX)} framework provides libraries for creating ML pipe"-lines for Data"=oriented, Model"=oriented and Operations Stages, as well as the \emph{ML Metadata (MLMD)} library for metadata management and version control \cite{Baylor17a,MLDM22a,MLDM22b}.
Embedded in this technical ecosystem, \emph{Kubeflow} enables coordinated deployments of workflows on Kubernetes clusters \cite{Kubeflow22a,Kubeflow22b}.

Although providing a platform for distributed in"=memory ML and predictive analysis on big data, \emph{H2O}'s \cite{H2O22a,H2O22b} ML lifecycle integration and ML artifact management capabilities are mostly comparable. Additionally, H2O aims at easy productionalization of ML models in enterprise environments.

\emph{Neptune} \cite{Neptune22a}, \emph{FBLearner Flow \& Predictor} \cite{Andrews16a,Dunn16a}, \emph{MLCask} \cite{Luo21a}, and \emph{Disdat}~\cite{Yocum19a,Disdat22a} additionally support software artifacts and further promote reproducibility.
Neptune provides dataset, model, and metadata artifact storage for pipelines, experiment tracking, and workspaces for coordination of projects and collaborating users.
FBLearner Flow \& Predictor is Facebook's proprietary ML pipeline development and processing system. It comprises Flow, a DAG"=based pipeline management system that facilitates experimentation, training, and comparison of models, and Predictor, an online inference framework based on an own model format.
MLCask is a pipeline"=oriented AMS. It builds upon Git"=inspired versioning of pipeline components with non"=linear version control semantics for collaborative environments.
Disdat manages ML pipelines and related ML artifacts by building upon two core abstractions: bundles and contexts. A bundle is a versioned, typed, immutable collection of data and files. A context is a view abstraction gathering a sharable set of bundles and assisting with managing bundles across multiple locations such as local and cloud storage environments.

\paragraph{Experiment Management}

Systems and platforms of this class aim to achieve comparability and reproducibility of exploratory ML experiments for model development, training, and optimization. They typically complement model training frameworks and AMSs of the previous class, since the results often serve as a starting point for subsequent pipeline creation and execution. Because of this, there is typically no or only limited support for the Model Operations Stage, but support for metadata and especially software artifacts.

The \emph{Deep-water Framework (DWF)} \cite{Ferenc20a,DWF22a} has a basic set of functionality that enables tracking of experiments and training runs, involved artifacts' identifiers, as well as configurations and performance metrics. DWF supports only a predefined set of models provided by TensorFlow and scikit"=learn and no storage or versioning of trained models.
\emph{StudioML} \cite{StudioML22a,StudioML22b} additionally captures model artifacts without the necessity of modifying experiment code.

Focusing on reproducibility for the Model Development and Model Operations Stages, \emph{ML"-Cube}~\cite{MLCube22a,MLCube22b} and \emph{MLPM}~\cite{Yao20a,MLPM22a} both provide model packaging capabilities:
MLCube is a library for packaging ML tasks and models, which enables sharing and consistent reproduction of models, experiments, and benchmarks.
MLPM enables users to adopt existing ML algorithms and libraries, resolving dependencies, and deploying as HTTP services.

An even higher degree of reproducibility is provided by \emph{Guild AI} \cite{GuildAI22a,GuildAI22b}, \emph{Datmo} \cite{Datmo22a,Datmo22b}, \emph{Deepkit} \cite{Deepkit22a,Deepkit22b}, and \emph{Keepsake} \cite{Keepsake22a,Keepsake22b}, which additionally capture any necessary software including source code, dependencies, execution environment, and logs. \emph{Runway} \cite{Tsay18a}, \emph{Sacred} \cite{Greff17a,Sacred22a}, and \emph{Weights \& Biases (W\&B)} \cite{WnB20a,WnB22a} furthermore provide capabilities for the management of data"=related artifacts.

\emph{Apache Submarine} \cite{Chen21a,Submarine22a,Submarine22b} and \emph{Determined} \cite{Determined22a,Determined22b} both provide functional interfaces and integration for popular ML training, experimentation, artifact management frameworks (\eg{} TensorFlow, PyTorch, MLflow, and TensorBoard) and Python SDKs for different stages of model development without requiring extra infrastructure knowledge for orchestration. Submarine supports both on"=premise clusters managed by Kubernetes or YARN, and clouds to ensure portability and resource"=efficiency.

\paragraph{Model Management}

AMSs of this class treat models and model metadata as central abstractions. While typically limited integration with the Data"=oriented Stage and capabilities for managing data"=related artifacts are provided, the focus is on the lifecycle steps for model development and operations as well as the management of models and their metadata. Although there are some functionality"=related intersections with the class of experiment management systems, AMSs of this class often provide support for the Operations Stage.

As two of the first model"=oriented AMSs, \emph{ModelDB} \cite{Vartak16a,Vartak17a,Vartak18a,Verta22a,ModelDB22a} and \emph{ModelHub} \cite{Miao16a,Miao17a,Miao17b} both focus on supporting model development, deployment, and monitoring.
While Model"-DB versions models and their metadata in a relational database, Model"-Hub incorporates an ML artifact versioning system enriching and extending Git, and a read-""optimized parameter archival storage that minimizes storage footprint using deltas and accelerates query workloads with minimal loss of accuracy.

\emph{ModelKB} is an AMS with a focus on model management, experimentation, deployment, and monitoring \cite{Gharibi19a,Gharibi19b,Gharibi21a}. It uses custom callbacks in native ML frameworks to collect metadata about each experiment and automatically generates source code for deployment, sharing, and reproducibility.

The \emph{MMP} is a model management platform tailored to Industry 4.0 environments by associating ML models with business and domain metadata \cite{Weber20a}. It provides a model metadata extractor, a model registry, and a context manager to store model metadata in a central metadata store.

Compared to the previously discussed systems, \emph{MISTIQUE} is specialized for the storage and management of model intermediates (\eg{} input data, learned hidden representations) to accelerate model evaluation, performance diagnosis, and interpretability \cite{Vartak18b}. It decides for each diagnostic query whether to re-run the model or to read a previously stored intermediate and reduces storage footprint of model intermediates with storage optimizations such as quantization, summarization, and data de"=duplication.

Motivated by fragmented ML workflows which require juggling between different programming par"-a"-digms and software systems, ML model training and inference algorithms as well as model management capabilities are increasingly integrated directly into DBMSs. \emph{Sql4ml} enables expressing supervised ML models in SQL and translating them into Python code for training in TensorFlow \cite{Makrynioti19a,sql4ml22a}.

\emph{Vertica"=ML} is an ML extension on top of the distributed and parallelized RDBMS \emph{Vertica Analytic Database} \cite{Fard20a,VerticaML22a,Lamb12a}, which aims for eliminating the transfer of big volumes of data, avoiding the maintenance of a separate analytical system, and addressing concerns of data security and provenance by combining a full"=fledged DBMS with the scalability and performance of in"=database ML algorithms.

\emph{ML"-Model"-CI}~\cite{Zhang20a,MLModelCI22a,MLModelCI22b}, \emph{Model"-CI"=e}~\cite{Huang21a}, \emph{Clipper}~\cite{Crankshaw17a,Clipper22a}, \emph{Rafiki} \cite{Wang18a,Rafiki22a}, and \emph{Overton}~\cite{Re20a} are AMSs focusing on the Model Operations stage.
ML"-Model"-CI is an MLOps platform for the automated deployment of pre"=trained ML models and online model serving. Profiling under different settings (\eg{} batch size and hardware) provides guidelines for balancing the trade"=off between performance and cost. For the deployment to cloud environments, ML"-Model"-CI uses Docker.
Model"-CI"=e is a plugin system for continual learning and deployment, enabling model updating and validation without model serving engine adaption.
Clipper is a general"=purpose low"=latency model serving system. By exploiting caching, batching, and adaptive model selection techniques, Clipper reduces prediction latency and improves prediction throughput, accuracy, and robustness without modifying underlying ML frameworks.
Rafiki provides a model training service supporting distributed hyperparameter tuning and a model inference service with online model ensembling that is amenable to the trade"=off between latency and accuracy.
Overton is an AMS focusing on building, deploying, and monitoring production models. It aims to support ML engineers in maintaining and improving model quality in the face of changes to the input distribution and new production features.

\emph{CMS} is a container-based continuous learning and serving platform designed for industrial monitoring and analysis use cases \cite{Li20a}. Its primary goal is to simplify and automate the process of model generation, deployment, and switching. Building on top of the Kubernetes management platform Rancher, CMS provides resource"=efficient orchestration of model training tasks and seamless model switching and serving without interruption of online operations and with minimal human interference.

\emph{ModelHub.AI} is a community-driven platform for the dissemination of deep learning models \cite{Hosny19a,ModelHubAI22a,ModelHubAI22b}. It is founded on a container"=based software engine that provides a standard template for models and exposes interfaces for model"=specific functions as well as data pre- and post"=processing. ModelHub.AI is domain-, data-, and framework"=agnostic, catering to different workflows and contributor's preferences.

\paragraph{Dataset \& Feature Management}

Complementary to the previously described class, this class focuses on support for the Data"=oriented Stage by providing dataset, label, and feature storage and management capabilities as well as functionality and interfaces for data (pre)processing, feature selection and engineering, and provenance tasks.

\emph{MLdp} is Apple's platform for managing ML data artifacts \cite{Agrawal19a}. It provides a minimalist and flexible data model for integrating different varieties of data, a hybrid storage approach for large volumes of raw data and high concurrent updates on volatile data, version and dependency management, data provenance, and integration with major ML frameworks.

In comparison, \emph{ExDra} provides an infrastructure for data acquisition, integration, and preprocessing from federated and heterogeneous raw data sources \cite{Baunsgaard21a}. It uses System"-DS for federated linear algebra programs, parameter servers, and data processing pipelines. Trained models and their provenance are stored in a model management database.

\emph{Pachyderm} is a data pipeline management platform \cite{Pachyderm22a,Pachyderm22b}. It provides automated data versioning, containerized pipeline execution, as well as immutable data lineage and provenance. Also motivated by enabling explainability, \emph{Prov"-Lake} is a data management system capable of capturing, integrating, and querying data across multiple distributed services, programs, databases, stores, and computational workflows by leveraging provenance data \cite{Souza19a,Souza19b,Souza21a,ProvLake22a,ProvLake22b}.

Unlike the previous systems, \emph{Data Provenance Library} \cite{Chapman20a,DataProvenance22a} and \emph{Shuffler} \cite{Toropov19a,Shuffler22a} operate at the library level.
Data Provenance Library is a Python library for capturing and querying fine"=grained provenance of data preprocessing pipelines. It is based on a formal model comprising data reduction, augmentation and transformation operators, as well as a MongoDB database as provenance store.
Shuffler is a toolbox for data preparation workflows of computer vision tasks. It employs relational databases and SQL for storing and manipulating annotations.

\emph{Feast} is a feature store for managing and serving ML features to models in production \cite{Feast22a,Feast22b}. Feast aims for enabling DevOps"=like practices for the lifecycle of features. As a single source of truth, Feast serves feature data either from a low"=latency online store for real"=time prediction, or from an offline store for scale"=out batch scoring or model training.

\subsection{Discussion Along Criteria}
\label{sec:assessment:criteriadiscussion}

This section discusses the assessment results comparatively along the criteria and subcriteria.

\paragraph{Lifecycle Integration}

Requirements engineering in the context of ML is a young field of research \cite{Ahmad21a,Villamizar21a,Vogelsang19a}. Many of the methods and approaches known from traditional software engineering have yet to be adapted for ML systems \cite{Arpteg18a,Belani19a,Lwakatare19a}. Additionally, the specification of non"=trivial requirements often necessitates domain expert knowledge. Consequently, the tool support is still poor and requirements engineering functionality is not yet covered by the assessed AMSs.

In contrast, many of the systems and platforms studied do at least provide partial functional support or interfaces for the Data"=oriented Stage: While most systems and platforms lack support for the Data Collection step, probably due to the individuality of data type, volume, sources, and collection approaches, many integrate functionality or provide functional interfaces for at least one of the Data Preparation~\& Cleaning, Data Labeling, and Feature Engineering~\& Selection steps (ca.~75\,\%).

Systems and platforms for lifecycle management offer the widest range of functions: Cloud platforms with integrated artifact management often provide their own tools with a graphical UI. Open"=source systems such as ClearML or MLflow, on the other hand, offer interfaces for integrating user code, which can then be used within pipelines for automation. In particular, AMSs with a narrow focus stand out here: For example, Feast and Hopsworks provide feature stores that are designed specifically for ML feature selection, storage, processing, and distribution.

A large proportion of the systems and platforms assessed provide wide or complete support for the Model"=oriented Stage (86\,\%), including both integrated functionality and interfaces to typical ML frameworks (TensorFlow, PyTorch, etc.), which provide functional support for model building, training, evaluation, and optimization. Nevertheless, some systems deviate from this due to their goals: for example, MMP and Vertica"=ML do not support \emph{Model Design} but have an integrated set of ML models.

The functional support for deployment and monitoring of models (\emph{Operations Stage}) is quite heterogeneous: 27 of \totalnumsys{} (ca.~42\,\%) provide full support and 14 of \totalnumsys{} (ca.~22\,\%) partial support. A majority of the lifecycle management systems and platforms provide functionality for deploying models as web services (\eg{} via REST interfaces) as well as continuous collection and monitoring of performance and quality metrics. In 6~out of 13~systems from the pipeline management class this is also the case, 3~further systems only provide support for deploying model serving environments. In the remaining 3~classes, the support is much lower due to the objective of the corresponding systems and platforms.

\paragraph{Artifact Support}

With more than 92\,\% on average, a large proportion of the systems and platforms takes model artifacts into account. The proportion for data"=related artifacts is lower at ca. 80\,\% (support for at least one type). Model"=specific metadata such as hyperparameters or metrics are collected and processed by more than 80\,\% of the systems. Experiment/""project metadata and pipelines are supported by ca.~67\,\% \resp{} 55\,\% of all systems. Only every second system takes software artifacts into account. The kind of support is strongly dependent on the objective and system class and is very heterogeneous across all systems and platforms; the relationship to the supported operations must be considered.

\paragraph{Operations}

With more than 90\,\%, the majority of the systems and platforms offers artifact capturing and logging functionality. Depending on the use of repositories and comparable techniques, only less than a half enable snapshots and intermediate states of artifacts to be checked in and rolled out again. Also, over 90\,\% provide operations to query and retrieve stored artifacts. More than two"=thirds (ca.~67\,\%) have comparison functionality, ca.~52\,\% provide artifact lineage, and only ca.~17\,\% offer provenance functionality. Visualization operations are present in ca.~62\,\% of the systems. While the presence of typical management operations such as modify, delete, and execute \& run is quite common, deployment operations are only present in about half of all systems. While the platforms and systems of the lifecycle, experiment, and model management classes mostly provide complete functionality for export and import as well as sharing with other collaborators, these functions are less prominent or not available at all for the other classes.

\paragraph{Collection \& Storage}

The collection of artifacts can either require explicitly added instructions (subcriterion \emph{Intrusive}), such as Python functions or callbacks, or be (semi-)""automatic (subcriterion \emph{Non"=intrusive}). While exactly half of the systems provide both intrusive and non"=intrusive collection of artifacts, primarily of the lifecycle, pipeline, and experiment management classes, almost two"=thirds of all systems (ca.~64\,\%) at least support automatic collection.

The types of storage used are highly dependent on the goals and focus of a system or platform. For example, lifecycle management systems provide an appropriate type of storage for each type of artifact (see §\,\ref{sec:criteria} for related discussion). It is also recognizable for the other system classes that the supported storage types are related to the supported artifacts themselves. Exceptions to this are systems and platforms such as Velox or MMP, which are tailored to specific domains and for this reason only support limited number of dataset and model types, as well as MISTIQUE or sql4ml, in which deep learning models are storage based on individual data models or supervised ML models in relational table structures.

In total 27~of the \totalnumsys{}~systems and platforms complementarity support both the complete versioning of a project or lifecycle state including any artifacts and the versioning of individual artifact snapshots. About half of all systems and platforms support at least repositories for versioning, whereby in addition to the general"=purpose version control systems, variants tailored specifically to ML are increasingly being used, which, for example, perform effective model versioning using deltas and provide special commands for model, dataset, pipeline, and experiment comparison and lineage information, as demonstrated by ModelHub's dlv and DVC.

\paragraph{Interfaces \& Integration}

Overall, many of the examined systems and platforms provide a wide range of interfaces and integration. While Python SDKs or REST APIs are provided by almost 9~out of 10~systems and platforms (ca.~88\,\%), CLI tools and web UIs are available in over two"=thirds (ca.~66\,\% \resp{} ca.~73\,\%). Here, especially the lifecycle management and experiment management classes stand out.

The programming language support (ca.~94\,\%)~-- primarily Python as the data science quasi"=standard~--, the integration for ML and data science frameworks (ca.~92\,\%) as well as the direct or indirect support for interactive and collaborative notebooks (ca.~92\,\%) is pronounced across all classes of systems and platforms. In particular, because of their focus, the systems and platforms of the lifecycle, pipeline, and experiment management classes have the highest degree of functional integration, which is related to the first category Lifecycle Integration.

\paragraph{Operation \& Licensing}

The capabilities for system and platform operation are highly dependent on the corresponding software architecture. 26~systems allow only one, 14~two, and 24~all three modes of operation. Over half of all systems can be used locally (ca.~54\,\%), either as a library, local server application, Docker container, or locally executable Kubernetes variant such as Minikube or Kind. Two"=thirds of the systems (ca.~66\,\%) can be deployed on"=premise (\eg{} on a dedicated server or cluster) and three"=quarters are capable of running in a cloud (ca.~75\,\%).

Among the systems and platforms studied, a total of 35~systems have some kind of free license, typically with source code freely available, and the remaining 32 have a non"=free, proprietary license.


\section{Conclusion}
\label{sec:concl}

This paper discusses system support for ML artifact management as an essential building block to achieve comparability, reproducibility, and traceability of artifacts created and used within the ML lifecycle. Objectives, fields of application, and functional ranges are heterogeneous and the selection of AMSs is quite difficult. Based on a systematic literature review, we derive functional and non"=functional criteria that enable the systematic assessment of AMSs. Using the criteria, we assess and discuss a comprehensive selection of \totalnumsys{}~systems and platforms from academia and industry.

As complementary and future work, we aim to investigate system support for automating ML tasks, \eg{} AutoML techniques such as automated hyperparameter optimization, neural architecture search, and meta"=learning, as well as for establishing ML"=related security properties, \eg{} techniques for hardening against and preventing model exploratory, data poisoning, and evasion attacks.

\begin{acks}
	This work was partially funded by the Thuringian Ministry of Economic Affairs, Science and Digital Society (grant 5575/10-3).
\end{acks}


\end{document}